\documentclass{article}

\usepackage{graphicx}      
\usepackage{cite}     
\usepackage{amsmath,amssymb,amsfonts}
\newcommand{\Vl}{{V\hspace{-.3mm}l}}
\usepackage{authblk}

\date{}

\title{Closed-loop robust control of long-term diabetes progression\\ via physical activity management\footnote{This manuscript is the arxiv version of the work by the author in \cite{Rocond2025control} }}

\author{
    Pierluigi Francesco De Paola$^{1,2,3}$, Alessandro Borri$^{*,3}$,\\ Fabrizio Dabbene$^{2}$,
    Pasquale Palumbo$^{4}$,
    Alessia Paglialonga$^{2}$
}

\affil{
\small 
    $^1$Politecnico di Bari, Bari, Italy\\
    $^2$Consiglio Nazionale delle Ricerche, Istituto di Elettronica e di Ingegneria dell’Informazione e delle
Telecomunicazioni (CNR-IEIIT), Turin, Italy\\
    $^3$Consiglio Nazionale delle Ricerche, Istituto di  Analisi dei Sistemi ed Informatica (CNR-IASI), Rome, Italy\\
    $^4$University of Milano-Bicocca, Department of Biotechnologies and Biosciences, Milan, Italy\\
    $^*$Corresponding author: alessandro.borri@iasi.cnr.it
}
\begin{document}

\maketitle

\begin{abstract}                

Large clinical evidence acknowledges the crucial role played by physical activity in delaying the progression of type-2 diabetes. However, the literature lacks control approaches that leverage exercise for type-2 diabetes control and more in general lacks a quantitative assessment of medical guidelines on the recommended amount of physical activity to be performed, mainly due to the absence of mathematical models that suitably 
estimate its benefits on diabetes progression. 
In this work, in order to provide a control-theoretical formulation of the exercise, we design a feedback law in terms of recommended physical activity, following a model predictive control
approach, based on a widespread compact diabetes progression model, suitably modified to properly account for the long-term effect of the exercise.  
Moreover we illustrate how the proposed approach proves to show reliable robustness properties with respect to initial conditions and parameter perturbations, which may be used to reflect inter-patient variability. Results are encouraging in view of the validation of the control law on comprehensive high-dimensional models of diabetes progression, with the aim of translating the prediction of the controller into reasonable recommendations and to quantitatively support medical decision-making. 
\end{abstract}


\section{Introduction}
\label{sec:introduction}

{Type} 2 diabetes (T2D) is a chronic disease forecasted to become a concerning challenge worldwide, in light of its severe complications and its rising {incidence}, with non-negligible social and economic implications on healthcare systems \cite{ZimmetEtAl2001}.
T2D is characterized by impairments in the glucose-insulin regulation mechanisms due to defects in insulin release by beta cells and {decreased} effectiveness of insulin itself in lowering glucose levels \cite{NDDG1979}.
The emerging interest for the glucose-insulin regulation mechanisms promoted in the last years the development of a successful research line named ``Artificial Pancreas",  focusing on  short-term diabetes control, with a huge variety of control applications, ranging from robust control \cite{KovacsEtAl2013}, symbolic control \cite{BorriEtAl2021}, model predictive control \cite{IncremonaEtAl2018,MessoriEtAl2018}, and nonlinear control \cite{BorriEtAl2017}. In this context, the availability of mathematical models  describing the glucose-insulin feedback loop plays a crucial role, as it allows the design and the development of model-based control techniques, accounting for the physiological meaning of the variables and parameters involved, thus proving to more reliable and effective for the control of biological systems with respect to the model-less techniques \cite{BorriEtAl2024}.
Within the framework of the Artificial Pancreas, all the approaches in the literature have been focusing on T2D control via insulin administration, that can be considered the last mile for T2D control, when the course of disease has become not reversible. 
As known from clinical evidence, T2D {arises} because of unhealthy lifestyle, promoting  over years the onset of insulin resistance and beta-cell degradation, in turn causing hyperglycemia and severe related issues.
In light of its aetiology, T2D can be prevented or significantly slowed down through lifestyle interventions, for example through physical activity, as well acknowledged in the clinical literature, see e.g. \cite{HemmingsenEtAl08,DiabetesGroup2002,WhelanEtAl2022} and references therein.
Nevertheless, the lack {of} mathematical models suitably describing the physiological mechanisms involving the effect of physical activity on diabetes course led to an {important} gap in the research line involving glucose control, with the results that no model-based control techniques so far have leveraged physical activity management for glucose control and diabetes prevention in the long term.
Recently, {the authors of this work} have developed {what appears to be} the first model in the literature \cite{DePaolaEtAl}, with its extension \cite{de_paola_novel_2024}, able to explain {-- and quantify --} the cumulative, long-term effects of physical activity in diabetes prevention. {This is obtained} by formalizing the role of the protein interleukin-6 (IL-6), whose release under regular exercise contributes in preserving beta-cell mass and function in the long term.
Hence, the model in \cite{DePaolaEtAl} and \cite{de_paola_novel_2024} paves the way to a novel and original model-based approach for diabetes control, {properly} leveraging the beneficial effects of physical exercise. 
Indeed, we have shown in a preliminary work \cite{DePaolaMIE2025} how it is possible to derive a compact, control-driven representation of the effect of the exercise starting from the formulation of the benefits of physical activity in our model~\cite{DePaolaEtAl},\cite{de_paola_novel_2024}. Specifically, in \cite{DePaolaMIE2025} we leverage a novel control-oriented formulation of the exercise effect to design a control law by a {model predictive control (MPC)} approach \cite{Allgower2012,Camacho2013book} on a compact model of diabetes progression, the model by ~\cite{ToppEtAl2000}, acknowledged to be a landmark in the field of diabetes modeling and suitably adapted to account for the effects of physical activity. In this work we build on preliminary findings in \cite{DePaolaMIE2025} to robustly verify the control-based approach and the contribution of this proposal is multi-fold: first, we compare the MPC formulation with a feed-forward control, showing the advantage of the MPC control in terms of total control effort; then, we consider the robust scenario and highlight that the proposed control approach proves reliable in presence of initial conditions and parameters perturbations for the considered model.
Hence, we provide in this work a novel model-based approach for long-term diabetes control via physical activity management and we show how our results are encouraging and deserve to be validated on a higher dimensional model \cite{DePaolaEtAl}, with the aim of providing a quantitative assessment to the general, experience-driven, medical advice on exercise programs for T2D prevention~\cite{bull2020world}.
The remainder of the paper is organized as follows: Section II illustrates 
the methodology followed for the definition of a control-driven formulation for glucose control via physical activity. Section III defines the design of the control law by means of an MPC strategy, Section IV shows the results we obtained in the different stages of implementation and, finally, in Section V we summarize our results and propose further extensions of the present work.
\section{A CONTROL-DRIVEN FORMULATION FOR PHYSICAL ACTIVITY}
\label{sec:SimulationResults}
The model here exploited to design the control algorithm is {a  modification to the model by \cite{ToppEtAl2000}}, and describes the evolution of glucose/insulin regulation mechanism in the a long period (months/years):
\begin{subequations}
	\begin{align}
 &\dot G = R_0- \left(E_{g0}+S_I I\right) G, \\
		&\dot I =\beta\sigma\frac{G^2}{\alpha+G^2}- kI,\\
        &\dot \beta = {(\bar{P}-\bar{A})}\beta,\\        
		&{\dot S_I  = -c(S_I-S_{I\!,\text{\it target}})}\cdot(1-\zeta_{si}\frac{\Vl}{\ {k_{n,si}}+\Vl}), \label{eq:SI}\\
  &\dot{\Vl} = \frac{SR}{K_{IL6}}\cdot  {u}  - k_s \Vl,\label{Vl} 
    \end{align}\label{eq:model}
\end{subequations}
where we omit time dependencies, with the state variables defined as follows:
\begin{itemize}
    \item[-] $G$ is the plasma glucose concentration [mg/dl];
	\item[-] $I$ is the serum insulin concentration [$\mu$U/ml];
	\item[-] $\beta$ is the beta cell mass [mg]; 
	\item[-] $S_I$ is the insulin sensitivity [ml/$\mu$U/d];
    \item[-] {$\Vl$ 	accounts for  the integral effect of IL-6 released during exercise sessions [(pg/ml)min].}
 \end{itemize}
 
{It should be observed that} we inherit the state variable $\Vl$ from our previous works~\cite{DePaolaEtAl},\cite{de_paola_novel_2024}, as it reports the effects of the physical activity on insulin sensitivity and beta-cells dynamics.
{In particular, we have}
\begin{subequations}
\begin{align} 
    \bar{P} &= P(G) \cdot \psi_1(Vl),\qquad \psi_1(Vl)=\left(1 + \frac{\zeta_{p}\cdot Vl^2}{k_{p}^2 + Vl^2}\right),\\
    \bar{A} &= A(G) \cdot \psi_2(Vl),\qquad \psi_2(Vl)=\left(1 - \frac{\zeta_{a} \cdot Vl^2}{k_{a}^2 + Vl^2}\right), \\
    {P}(G) &=(r_{1r} \cdot G - r_{2r}  \cdot G^2), \\
    {A}(G)&= (d_0 - r_{1a} \cdot G + r_{2a} \cdot G^2).
\end{align}
\end{subequations}

Glucose/insulin, beta-cells and insulin sensitivity dynamics are inherited from \cite{ToppEtAl2000}. In more detail, the beta-cell dynamics is modified with respect to the original work of \cite{ToppEtAl2000} were $\psi_1$ and $\psi_2$ where formally set to 1; we reformulated this equation to highlight the separate contribution of physical activity on beta-cell proliferation ${P}(G)$ and apoptosis ${A}(G)$ 
by means of Hill functions of the $\Vl$ state variable, similarly to our previous works \cite{DePaolaEtAl},\cite{de_paola_novel_2024}. Besides, with respect to \cite{ToppEtAl2000}, we introduce a general positive $S_{I\!,\text{\it target}}$ in  $S_I$ dynamics, as in our works \cite{DePaolaEtAl}, \cite{de_paola_novel_2024}, to avoid the unrealistic increase in beta-cell mass that characterizes the original model by \cite{ToppEtAl2000}.
Finally, we model the additional effect of exercise on insulin sensitivity, widely known in the literature \cite{bird2017update}, according to the formulation in \cite{de_paola_novel_2024}.
Specifically, the effect of exercise on $S_I$ is modeled by means of a Michaelis-Menten function of the $\Vl$ state variable to account for a larger increase of $S_I$  following the first weeks of training, with the benefits almost vanishing in the long term, according to the evidence in the literature \cite{bird2017update}.
Parameters of the model are listed in Table \ref{tab:Topp's_parameters_values_units_extended}.
\begin{table}
  \centering
  \resizebox{0.4\textwidth}{!}{
 \begin{tabular}{ccc}
    \hline
    \textbf{Parameters} & \textbf{Value} & \textbf{Unit} \\
    \hline
    $R_0$    &  864 &  mg/dl/d \\
    $E_{g0}$ & 1.44 & 1/d  \\
    $\sigma$ & 43.2 & $\mu$U/ml/d \\
    $\alpha$ & 20000 & mg$^2$/dl$^2$ \\
    $k$      &  432 & 1/d \\
    $d_0$    & 0.06 & 1/d \\
    $c$      & $0.05$ &  1/d \\
    $r_{1r}$ & $0.42*10^{-3}$ & dl/mg/d  \\
    $r_{2r}$ & $0.12*10^{-5}$ & dl/mg/d  \\
    $r_{1a}$ & $0.42*10^{-3}$ & dl/mg/d \\
    $r_{2a}$ & $0.12*10^{-5}$     &  dl/mg/d              \\
    $\zeta_{p}$& $10^{-4}$ & / \\
    $k_{p}$  & $10^{6}$  & ((pg/ml)min) \\
    $\zeta_{a}$ & $10^{-3}$ & / \\
    $k_{a}$  & $10^{6}$ & ((pg/ml)min) \\
    $S_{I\!,\text{\it target}}$ & 0.028 &  ml/$\mu$U/d \\ 
    $\zeta_{si}$ & 1.4  & / \\
    $k_{n,si}$  & $5*10^{6}$ & (pg/ml)min\\
    $SR$    & 0.045 & pg/ml/min \\
    $K_{IL_6}$  & 0.004 & 1/min \\
    $k_{s}$    &  $-\frac{log(0.8)}{(80640)}$ & 1/min \\    
    \hline
  \end{tabular}
  }
  \caption{Parameters of the control model with values and units}  \label{tab:Topp's_parameters_values_units_extended}
\end{table}

The model \eqref{eq:model} can be readily restated in nonlinear state-space form
\begin{equation}
    \dot x=f(x,u),
    \label{eq:model_bis}
\end{equation}
with 
\begin{align*}
x&=[ x_1 \quad x_2 \quad x_3 \quad x_4 \quad x_5]^\text{T}\\
&=[ G \quad I \quad \beta \quad S_I \quad Vl  ]^\text{T}\in\mathbb{R}_{\geq 0}^5.    
\end{align*}

{For what concerns}
the control input, it is defined starting from the original formulation in \cite{DePaolaEtAl}, where physical activity enters a set of fast dynamics that  are here neglected to simplify the model-based control scheme, given the aim to control T2D progression in the long term.
Indeed, in \cite{DePaolaEtAl} the control input coming from physical activity was defined as
a piecewise-constant input representing the exercise intensity,
\begin{equation}
u(t)=\begin{cases}
0 & t\in[0,T-\delta) \quad \mod T\\
\bar{u} & t\in[T-\delta,T) \quad \mod T,
\end{cases}
\label{eq:PWC}
\end{equation}
where $\delta$ is the duration of the exercise sessions,  T is the period of the training and $\bar{u}$  is the activity level. Hence, the exercise program is described by the triplet $(\bar{u},\delta,T)$.

In this work, with the aim of deriving a control-theoretical formulation of the effect of the exercise on diabetes progression, similarly to what described in \cite{DePaolaMIE2025} we rely on an equivalent constant input $u_{eq}$ - representing the average effect of the physical activity on $Vl$:
\begin{align}
   u(t)=u_{eq} = \frac{\bar{u}\cdot\delta + 0\cdot (T-\delta)}{T} = \frac{\bar{u}\cdot \delta}{T}, \qquad t\in[0,T]. \label{eq_ueq}
\end{align}

According to \eqref{eq_ueq}, the equivalent input $u_{eq}$ represents the time average of the exercise intensity over the period~$T$. 
The $Vl$ dynamics comes from the equations of \cite{DePaolaEtAl} when the control input $u$ is supposed fixed to its average value $u_{eq}$ and the fast dynamics of \cite{DePaolaEtAl} describing how physical activity effects propagate on the glucose/insulin dynamics are at their \textit{quasi-stationary} values.

The concept of equivalent input $u_{eq}$ allows to represent the contribution of physical activity in a compact form, encoding all the features related to the exercise program and accounting for all the information about the amount of exercise to be performed.
With the aim of designing a feedback law by means of an MPC controller, the role of $u_{eq}$ provides a great advantage, as it allows for a continuous formulation of the control with respect to the original impulsive representation encoded through repeated exercise sessions as in \cite{DePaolaEtAl},\cite{de_paola_novel_2024}.
The general information of the computed control can then be traslated into a duration control, by keeping fixed the exercise intensity $\bar{u}$ and the period $T$ of the exercise sessions, by means of the inverse map
\begin{align}
  \delta =  \frac{ u_{eq} \cdot T}{\bar{u}} \label{eq_inverse_map}.
\end{align}

\begin{figure}[!ht]
\hspace*{-0.5 cm}
\centering
{\includegraphics[scale=0.5]{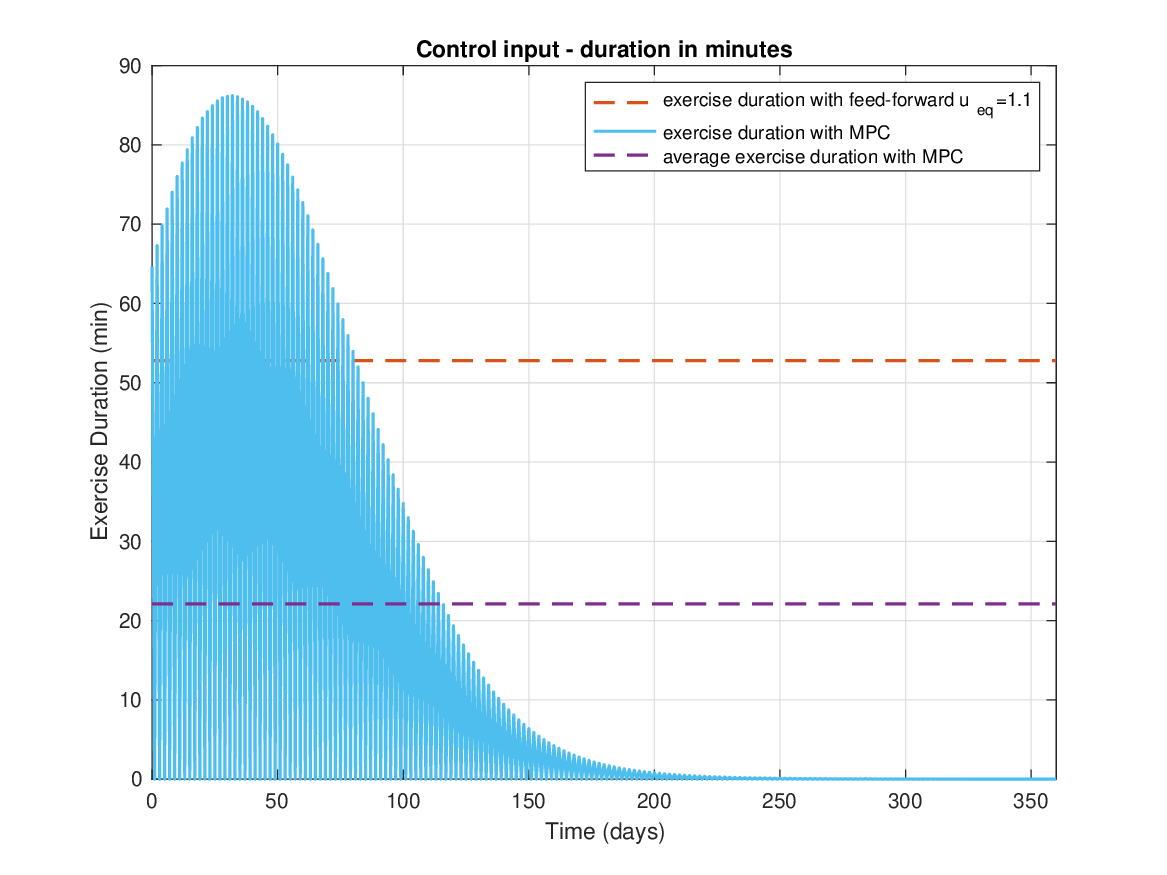}}
\caption{Recommended duration of single exercise sessions as a function of time computed by means of the inverse map \eqref{eq_inverse_map}, with the nominal parameters in Table \ref{tab:Topp's_parameters_values_units_extended}:  feedforward control case with $u_{eq}=1.1$ (dashed red line) vs MPC controlled case (blue stem lines and dashed purple line).} 
\label{fig:duration}
\end{figure}

In this way, the general exercise program encoded in $u_{eq}$ is converted into precise recommendations in terms of duration of the exercise to be performed. The recommendations are updated in accordance with the predictions of the controller (see e.g.  Fig. \ref{fig:duration} in Section \ref{sec:SimulationResults}), whose design is addressed next.
\section{MPC design }
In this section, we design a feedback law for glucose control via physical activity through a 
{model predictive control} approach  \cite{Camacho2013book}, which {represents one of } the most popular model-based solutions to the feedback control of real systems,  widely employed in industrial, technological and biomedical applications (see for example \cite{Afram2014,Faedo2017,MessoriEtAl2018,D'OrsiEtAl2024,BorriEtAl2024}), and which allows to properly take into account not only the model dynamics but also linear/nonlinear constraints on state and input variables. 
As typical in optimization approaches, MPC has the goal of achieving the best trade-off between conflicting goals, which in this context are:
\begin{itemize}
    \item the reduction of the diabetes progression, quantitatively evaluated in terms of glucose levels $G$;
    \item  the practical feasibility of the amount of exercise to be performed under a regular regime of physical activity, evaluated in terms of level of the equivalent control input $u_{eq}$.
\end{itemize} 

Consequently, we will next consider an integral cost function, with quadratic terms  defined in the running cost at each continuous time $t$, to take into account both performance (glycemic value $G$) and control effort ($u_{eq}$ level) during the whole evolution of the system.

In order to allow for a flexible tuning of the balance between performance and control effort, we also introduce a positive real penalty coefficient $\lambda$ weighing the control input. When $\lambda$
 tends to $0$, the control effort is scarcely accounted for in the optimization process, resulting in the trivial solution that the maximum allowed amount of physical activity level is recommended: $u_{eq}=u_{eq}^{\max}$, which is not realistic. On the other hand, for $\lambda$ growing sufficiently large, the achieved glycemia marginally affects the cost function, which will be optimized in absence of physical exercise: $u_{eq}=0$.

 On top of the continuous-time formulation of the cost function, due to the continuous-time nature of the system \eqref{eq:model} that we want to control, we assume a sampled-data implementation of the feedback strategy, i.e. the equivalent  input $u(t)$ is implemented as a piecewise-constant signal, with hold interval equal to the period of the physical activity:
 \[
  u(t)= u(kT), \quad t \in [kT,(k+1)T), \quad k \in \mathbb{N}.
 \]
 
 The motivation for the discrete-time implementation is twofold: from a computational viewpoint, the optimization is performed on discrete time instants, which is much less demanding; on the other hand, since the equivalent input will be translated into a train of pulses, whose duration is modulated once per period, in agreement with the inverse map \eqref{eq_inverse_map}, it would be pointless to obtain a continuous-time evaluation of the equivalent input, which we may want  to update (at most) at every training session.

 Finally, the essential feature of MPC, which distinguishes it from classical optimal control, is 
 the receding-horizon implementation:  the optimization is solved based on a model prediction from the current state, considering a proper prediction horizon, that we assume to be equal to $N$ training periods (also called \textit{prediction window}), while the obtained control strategy is not applied for the whole duration of the prediction window but only for a shorter \textit{control window}, which we assume to be equal to only one period. After each control period,  the prediction horizon shifts onwards and the optimization  is repeated, so that the control law is updated at each period (before each session of physical activity).
 This allows to have a more frequent feedback from the controlled system with respect to open-loop optimal control, a  preferable choice in presence e.g. of parameter uncertainties and unmodeled dynamics, while preserving computational tractability with respect to closed-loop optimal control, which is real-time unfeasible in most applications. We refer the interested reader to  \cite{Allgower2012,Camacho2013book} for a deeper discussion on this topic.
 
More formally, the MPC problem finds the optimal control sequence $\{u^*_{eq,k}\}_{k\in\mathbb{N}}$, whose general component is defined as follows:

\begin{align}\label{eq:u_star}
    u^*_{eq,k} = &\arg\min_{u_{eq}\in[0,u_{eq}^{\max}]} \displaystyle\int_{kT}^{(k+N)T} (x_1(s)^2  +\lambda\cdot u_{eq}^2)\, ds \\
    \hspace{-3mm}\text{s.t.} & \quad \begin{aligned}[t]
        & \dot{x} = f(x(t),u(t)), \quad  t\geq 0, \\
        & u(t)= u^*_{eq,k}, \quad \qquad  t \in [kT,(k+1)T), \quad  k \in \mathbb{N}.\nonumber
    \end{aligned}
\end{align}
\vskip 3mm
 
Due to the high non-linearity of the model \eqref{eq:model}, the optimal solution \eqref{eq:u_star} cannot be obtained in closed form and is computed numerically. 

As already mentioned, the constraints account for the dynamics of the system    \eqref{eq:model_bis} and for the  bounds on $u_{eq}$. More in detail, we set $u_{eq}^{\max}=3$ as, when this maximal regime is translated into a duration recommendation via the inverse map \eqref{eq_inverse_map} (with moderate-to-vigorous exercise, i.e. $\bar{u}=60\%$, $T=2$ days), it results in an overall duration of exercise very close to 400 minutes/week, which seems to be a saturating threshold for the effects of exercise on diabetes progression \cite{boonpor2023dose}.

\begin{figure}[!ht]
\hspace*{-0.5 cm}
\centering
{\includegraphics[scale=0.5]{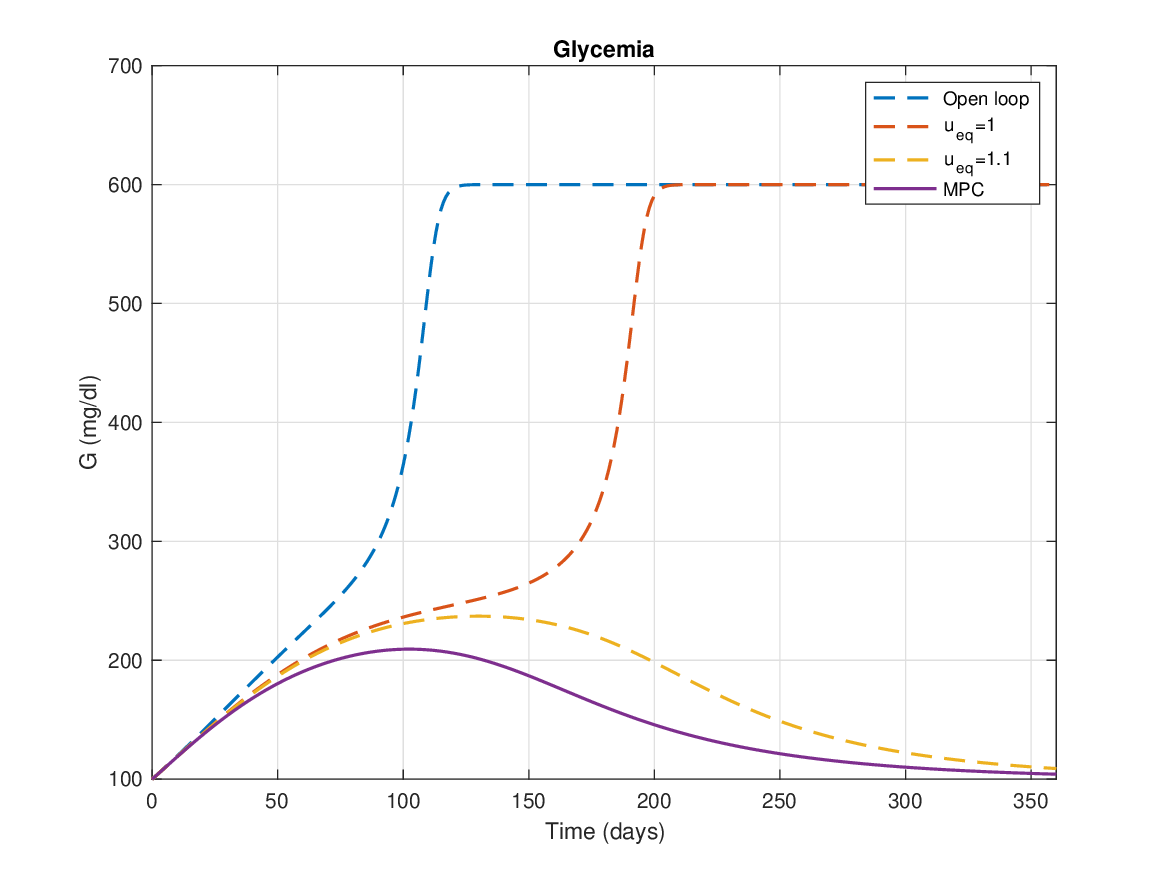}}
\caption{Basal glucose concentration as a function of time in the open-loop case (dashed blue line), in the feedforward control case (dashed red and yellow lines) and in the MPC controlled case (solid purple line) using the {nominal} parameters in Table \ref{tab:Topp's_parameters_values_units_extended}. }
\label{fig:G_basal}
\end{figure}

\begin{figure}[!ht]
\hspace*{-0.5 cm}
\centering
{\includegraphics[scale=0.5]{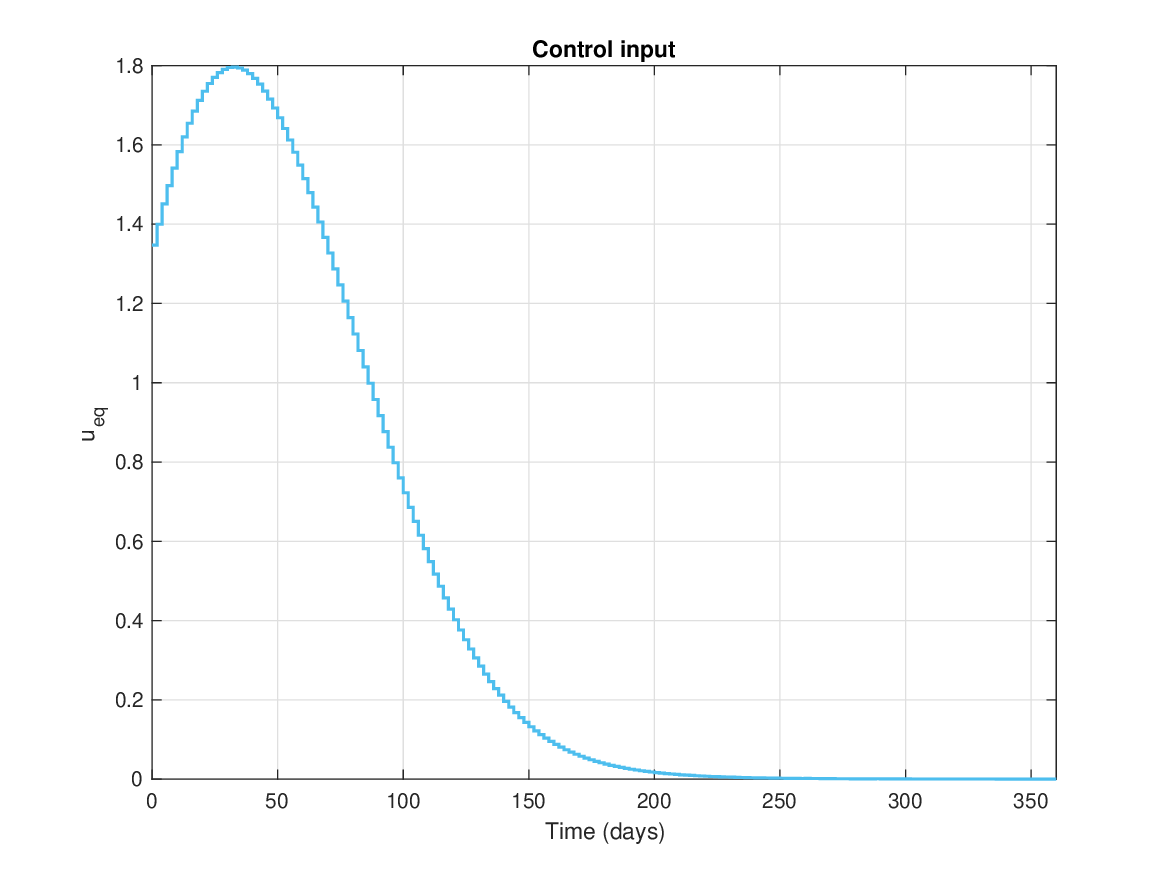}}
\caption{Equivalent control input $u_{eq}$ predicted by the MPC controller as a function of time, with the nominal parameters in Table \ref{tab:Topp's_parameters_values_units_extended}.}
\label{fig:u_eq}
\end{figure}

\begin{figure}[!ht]
\hspace*{-0.5 cm}
\centering
{\includegraphics[scale=0.5]{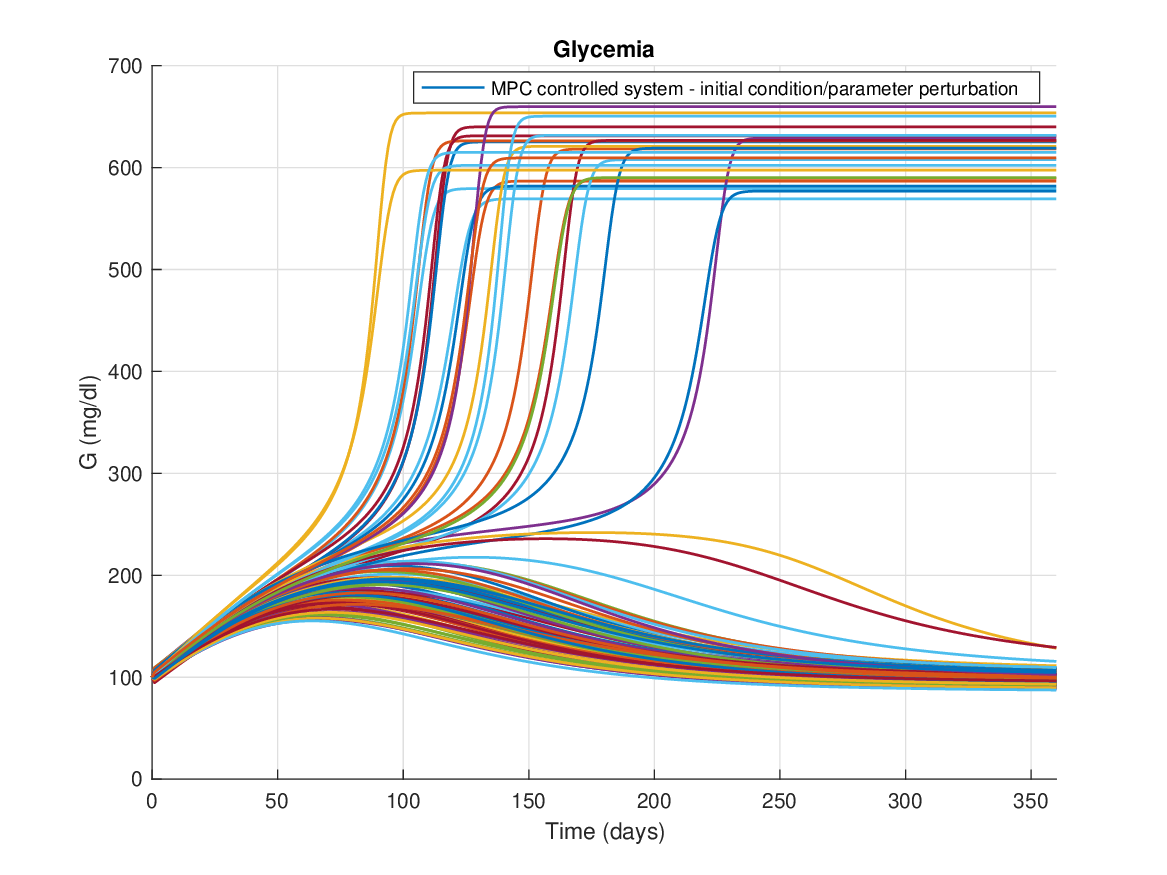}}
\caption{Basal glucose concentration as a function of time in the controlled case considering initial condition and parameter perturbations ($100$ simulations). }
\label{fig:G_MPC_100}
\end{figure}

\begin{figure}[!ht]
\hspace*{-0.5 cm}
\centering
{\includegraphics[scale=0.5]{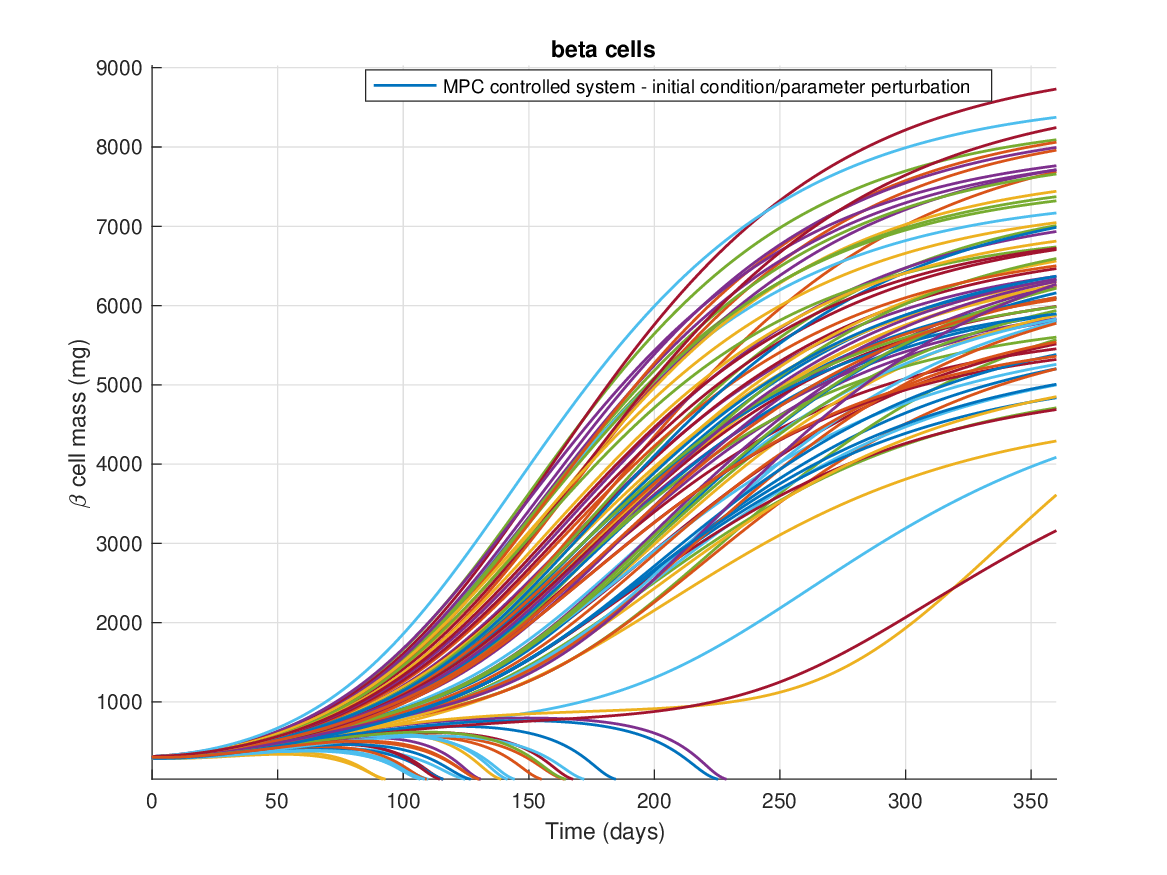}}
\caption{Beta-cell mass as a function of time in the controlled case considering initial condition and parameter perturbations ($100$ simulations).}
\label{fig:Beta_MPC_100}
\end{figure}

\section{Results}\label{sec:results}
We assume that the conditions predisposing to diabetes arise at $t=0$, with a severe insulin sensitivity decay according to Eq. \eqref{eq:SI} and Table \ref{tab:Topp's_parameters_values_units_extended}. 
Simulations were performed in MATLAB\textsuperscript{\textregistered} (version R2023a) (the gradient-based method \textit{fmincon} has been used for the MPC implementation), by assuming the  initial condition $x(0)=[100 \quad  10 \quad 300 \quad 0.72 \quad 0]^\text{T}$. The prediction window $N$ is set to $20$ to match the time constant of insulin sensitivity decay and parameter $\lambda$ is set equal to $60$, calibrated by simulation. The control horizon ($\theta$) is set equal to the period $T$ (2 days). Since diabetes progression according to the model by \cite{ToppEtAl2000} occurs tipically over a time span of one year, the same period is considered for our simulations ($\Theta=365$ days). 
We consider two simulation scenarios: 
\begin{itemize}
    \item [A)] simulations performed in the nominal setting of initial conditions and model parameters as described above;
    \item [B)] Monte Carlo simulations, accounting for initial conditions and model parameter perturbations.
\end{itemize}

 \subsection{Simulation with nominal initial conditions and model parameters}
 We simulate the model in the open-loop case, in the case of a feed-forward constant control $u_{eq}$ representing a constant overall exercise program and in the MPC controlled case. 
Fig. \ref{fig:G_basal} shows the trend of the basal glucose concentration. 
As it can be observed from the plot, in the open-loop case the severe diabetes progression promotes increasing glucose levels causing in turn beta cell degradation and irreversible disease course \cite{ToppEtAl2000}. As a result, the system reaches the hyperglycemic steady state of the Topp model at $G=600$ mg/dl. The same occurs with a feed-forward control, for the constant control values $u_{eq}$ considered ($u_{eq}=1)$. Although the exercise keeps constant throughout the simulation horizon, it is not capable of restoring normal values for glucose concentration and of avoiding an irreversible course for diabetes.
The feedforward control $u_{eq}=1.1$ is the minimum constant control input that allows the system to restore normoglycemia and stopping the progression of the disease, as shown in the plot.
In closed loop, the MPC recommendations, with the equivalent control input $u_{eq}$ updated according to the progression of the disease, allows the system to stop the course of diabetes and, in the long term, to reverse the progression and to keep the simulated subject safe, restoring normo-glycemic values and the steady-state at $G=100$ of the Topp model.
Fig. \ref{fig:u_eq} shows the plot of the equivalent control input $u_{eq,\text{MPC}}$ computed by the MPC, updated according to the stages of diabetes course. Specifically, the  control input assumes greater values 
in the first phases of diabetes course, thus subsequently decreasing and vanishing as the disease is slowed down and reversed. 
To show the advantage of the MPC implementation, we compare the total control effort in case of MPC controller and feedforward control. We define $u_{ff,\text{min}}:=1.1$ as the minimum value of the equivalent constant control $u_{eq}$ able to prevent diabetes progression and define the total control effort $\eta$ as 
\[
\eta(u):=\int_{0}^\Theta |u(t)|\,dt = \|u\|_{L^1_{[0,\Theta]}}.
\]
Computing $\eta(u_{ff,\text{min}})$ we get 
\[\eta(u_{ff,\text{min}})=u_{ff,\text{min}}\cdot \Theta
= 401.5, 
\]
while for the MPC controller we get
\[\eta(u_{eq,\text{MPC}})= \theta\cdot \sum_k u_{eq,k}=  168.1,  
\]
showing that the total control effort in the MPC implementation turns out to be significantly lower with respect to the case of feedforward control. 

While the quantity $\eta(u)$ defined above represents a cumulative exercise intensity over the considered time horizon, a more intuitive interpretation of the control effort is provided in Fig. \ref{fig:duration}, showing how the control law in terms of equivalent input $u_{eq}$, embedding a general information about the exercise program, is converted into a time-dependent precise recommendation on the duration of the exercise session (both for the MPC control and the feed-forward control $u_{ff,\text{min}}=1.1$). More in detail, we fix the exercise intensity  $\bar{u}=60\%$ (moderate-intensity exercise) and the period $T=2$ days, and we convert the optimal MPC input \eqref{eq:u_star} and the feed-forward control $u_{ff,\text{min}}=1.1$ into a precise recommendation on the duration of single exercise sessions to be undertaken through the map 
\begin{align}
  \delta_k^{\ast} =  \frac{ u_{eq,k}^{\ast} \cdot T}{\bar{u}} \label{eq_inverse_map_bis}.
\end{align}

Notice that the map in (\ref{eq_inverse_map_bis}) is the general formulation to account for the time-varying MPC control law. It is straightforward that, in the case of  feed-forward control, the equivalent input is constant, leading to the following minimum constant exercise duration required to prevent diabetes
\begin{align}
  \delta_{ff,\text{min}} =  \frac{ u_{ff,\text{min}} \cdot T}{\bar{u}} \label{eq_inverse_map_tris}.
\end{align}

From computations and Fig. \ref{fig:duration}, one readily obtains $ \delta_{ff,\text{min}}\simeq 52$ minutes, while the average MPC exercise duration is about $22$ minutes. 
 \\
Furthermore, on a weekly basis, 
the time-varying duration (in the MPC control) of the exercise sessions takes a maximum, in the early phases of diabetes progression, around $250$ min/week. These results are aligned with WHO general guidelines and with clinical evidence, 
suggesting to perform exercise beyond the general recommendation of 150 min/week \cite{bull2020world} to further reduce the risk of diabetes \cite{boonpor2023dose},\cite{li2008long}. Moreover, the MPC controlled system suggests higher doses of exercise in the initial phases and lower doses in the long term, in line with the fact that the benefits of physical activity may persist in the long term even after the discontinuation of the intervention \cite{uusitupa2003long}, 
confirming the primary role played by physical exercise in preserving beta cell mass from degradation in the early stages of diabetes course \cite{defronzo2011preservation}.
\subsection{Monte Carlo simulations}
In order to evaluate the robustness of the proposed method, the MPC-controlled model has been simulated ($100$ Monte Carlo simulations) assuming initial condition and parameter perturbations with relative variations with respect to the nominal values uniformly distributed over the interval $[-\phi, \phi]$, with $\phi=5\%$. Initial conditions and parameters perturbations are aimed at:
\begin{itemize}
    \item[ - ] simulating patients inter-variability;
	\item[ - ] accounting for the robustness of the control with respect to different simulated subjects.

\end{itemize}
Fig. \ref{fig:G_MPC_100} and Fig. \ref{fig:Beta_MPC_100} show the basal glucose concentration and beta-cell mass as 
a function of time for the controlled system in the $100$ simulations obtained by means of initial conditions and parameter perturbations as described above. The controlled system successfully prevents T2D in 75\% of simulations by restoring normoglycemia over the one-year simulation horizon, whereas irreversible diabetes progression is observed in only $25\%$ of the simulated cases, when physical exercise does not prove to be sufficient to prevent beta-cell mass from degradation due to  progression of the disease (Fig. \ref{fig:Beta_MPC_100}). 
It should be noted that the values of the beta cell mass at the end of the simulation is perfectly consistent with the mass of the human pancreas as discussed in \cite{DeGaetanoEtAl2008}.

\section{Conclusions and open issues}\label{sec:conclusions}

This work shows a proof of concept devoted to illustrate how to leverage physical activity and the different features (intensity, duration, period) of an exercise program to control diabetes progression. Results are consistent with the literature for what concerns the amount of the exercise to be performed according to the different stages of development of the disease, as already discussed. Moreover, results showed that the benefits of physical activity may vary with varying model parameters (that may relate to varying individual characteristics), in line with the well-known inter-individual variability of the benefits of lifestyle interventions for T2D prevention.
Our preliminary results are encouraging, however the study shows some limitations that need to be overcome in the future developments of the work. 
Specifically, we inherit some limitations of the model by \cite{ToppEtAl2000} that, despite being a pioneering work in the field of diabetes modeling, accounts for a time range of only one year, whereas a more realistic horizon for the development of the disease spans over two or more years \cite{DePaolaEtAl}, \cite{HaEtAl2016}. 
We expect to overcome these limitations implementing the concept of $u_{eq}$ on a quasi-steady state approximation (QSSA) of our extended model \cite{DePaolaEtAl,de_paola_novel_2024}, with the aim of designing the control law on the QSSA model and validating the control law on its extended version.

\bibliography{ifacconf}  
\bibliographystyle{ieeetr}
         
\end{document}